\documentclass[apl,amsmath,amssymb,amnsfonts,twocolumn,a4paper]{revtex4-1}
\usepackage[T1]{fontenc}
\usepackage{geometry}
\usepackage{graphicx}
\usepackage{color}
\usepackage[usenames,dvipsnames]{pstricks}
\usepackage{epsfig}
\usepackage{dcolumn}
\usepackage{booktabs}
\usepackage{pst-grad} 
\usepackage{pst-plot} 

\begin{document}
\title{Calculating thermal stability and attempt frequency of advanced recording structures without free parameters} 

\author{Christoph Vogler}
\email{christoph.vogler@tuwien.ac.at}
\affiliation{Institute of Solid State Physics, Vienna University of Technology, Wiedner Hauptstrasse 8-10, 1040 Vienna, Austria}

\author{Florian Bruckner}
\author{Dieter Suess}
\affiliation{Christian Doppler Laboratory for Advanced Magnetic Sensing and Materials, Institute for Solid State Physics, Vienna University of Technology, Wiedner Hauptstrasse 8-10, 1040 Vienna, Austria}

\author{Christoph Dellago}
\affiliation{University of Vienna, Faculty of Physics, Boltzmanngasse 5, 1090 Vienna, Austria}

\date{\today}

\begin{abstract}
Ensuring a permanent increase of magnetic storage densities is one of the main challenges in magnetic recording. Conventional approaches based on single phase grains are not suitable to achieve this goal, because their grain volume is limited due to the superparamagnetic limit. Grains with graded anisotropy are the most promising candidates to overcome this limit, providing magnetic memory bits with small volumes, low coercivity and high thermal stability at the same time. Combining micromagnetic simulations with forward flux sampling (FFS), a computational method for rare events that has been recently applied to the magnetic nanostructures, we have determined thermal escape rates and attempt frequencies of a graded media grain and two single phase grains of the same geometry. We find that graded anisotropy can increase the thermal stability of a grain by 12 orders of magnitudes from tens of milliseconds to centuries without  changing the coercive field.
\end{abstract}

\keywords{superparamagnetic limit; attempt frequency; graded media; recording trilemma; forward flux dampling; Nudged Elastic Band Method}

\maketitle 

Future magnetic recording devices will require high coercivity materials such as FePt in order to insure thermal stability at high storage densities. However, the requirements of small volume (for high densities), low coercivity (for good writeability) and high thermal stability can not be optimized at the same time using single phase grains, consisting of just one material.  While the magnetic reversal of small grains of high-coercivity materials requires fields that exceed the capability of state-of-the-art magnetic write heads, small grains of low-coercivity materials have insufficient thermal stability. Recently, a new type of grain was proposed that is easily writeable and, at the same time, thermally stable even for small grain sizes \cite{suess_multilayer_2006,suess_thermal_2008}. The grain consists of a stack of different materials with graded anisotropy, designed to overcome the superparamagnetic limit by reducing the coercive field while keeping a high barrier opposing spontaneous magnetization reversal. Calculations \cite{suess_multilayer_2006,suess_thermal_2008,dittrich_path_2002} carried out with the nudged elastic band (NEB) method \cite{henkelman_climbing_2000} confirm a high energy barrier $\Delta E_{\mathrm{b}}$ for the magnetization reversal of such graded media grains with small volumes and low coercivity.
Following the proposal of graded media grains, however a controversy has arisen \cite{dobin_attempt_frequency_2007} on the value of the attempt time $\tau_{0}$, which, together with the barrier height $\Delta E_{\mathrm{b}}$, determines the thermal relaxation time $\tau$ according to the Arrhenius-Nèel law
\begin{equation}
\label{eq:ArrheniusNeel}
  \tau=\tau_{0}e^{\frac{\Delta E_{\mathrm{b}}}{k_{\mathrm{B}}T}}.
\end{equation}
Here, $k_{\mathrm{B}}$ and $T$ are the Boltzmann constant and the temperature, respectively, and $\tau_0$ is the inverse of the attempt frequency, $f_0 = 1/\tau_0$. The relaxation time $\tau$ is the average time between spontaneous magnetization reversals and is a measure of the thermal stability of the grain. While one usually assumes that the attempt time $\tau_0$ is in the nanosecond range, it was suspected \cite{dobin_attempt_frequency_2007} that for graded media grains $\tau_0$ increases  over many orders of magnitude, thus lowering the total thermal stability of graded media grains significantly despite the high energy barrier $\Delta E_{\mathrm{b}}$. 

In this work we demonstrate, using computer simulations, that graded media grains are far from the superparamagnetic limit, providing high density storage devices with low coercivity and at the same time high thermal stability. We consider magnetic particles of various compositions and model their micromagnetic dynamics with a set of Langevin equations, which arise from the Landau-Lifshitz-Gilbert equation including thermal fluctuations,
\begin{eqnarray}
\label{eq:LLGThermal}
  \frac{\mathrm{d}\boldsymbol{\hat{m}}}{\mathrm{d}t}=&-&\frac{|\gamma|}{\left (1+\alpha^{2}  \right )}\left \{\boldsymbol{\hat{m}}\times \left (  \boldsymbol{H}_{\rm eff}[\boldsymbol{\hat{m}}]  + \boldsymbol{H}_{\rm th} \right )\right \}\nonumber \\
  &-& \frac{\alpha |\gamma|}{\left (1+\alpha^{2}  \right )}\left \{\boldsymbol{\hat{m}}\times \left [\boldsymbol{\hat{m}}\times\left (  \boldsymbol{H}_{\rm eff}[\boldsymbol{\hat{m}}] + \boldsymbol{H}_{\rm th} \right )  \right ]  \right \}.
\end{eqnarray}
In this equation, $\gamma$ is the reduced electron gyromagnetic ratio ($|\gamma| = |\gamma_{e}\cdot\mu_{0}| =2.213 \cdot 10^{5}$\,m/As), $\alpha$ is the damping parameter and $\boldsymbol{\hat{m}}$ is the space dependent magnetization of the magnetic particle normalized by its saturation magnetization. In the simulations, one configuration vector $\boldsymbol{\hat{m}}$ contains the spatial components of all $K$ computational nodes in the finite element model of the underlying particle and thus has $3K$ components. The effective magnetic field $\boldsymbol{H}_{\rm eff}[\boldsymbol{\hat{m}}]$ is obtained by taking the functional derivative of the total energy $E_{\mathrm{tot}}$ of the system with respect to the magnetization. We consider exchange, anisotropy, demagnetization and Zeeman energy contributions. $\boldsymbol{H}_{\rm th}$ describes a random thermal field with white noise properties, which accounts for the thermal activation of the system.

The thermally activated magnetization reversal of long-lived magnetic nanostructures is typically a rare event, i.e., the average time between magnetization reversals is much longer than the time scale for typical magnetization fluctuations of the system evolving according to Eq.~\ref{eq:LLGThermal}. In particular, the barrier crossing events take place quickly when they happen, but the waiting times between such events are long. Several techniques, including the string method \cite{e_string_2002}, transition path sampling \cite{dellago_transition_1998,bolhuis_transition_2002,van_erp_novel_2003}, and forward flux sampling (FFS) \cite{allen_sampling_2005,allen_simulating_2006}, have been put forward in the past decades to study the mechanism and the kinetics of such rare events in computer simulations. Here, we  compute relaxation times $\tau$ and attempt times $\tau_0$ of magnetic grains using FFS, which was recently shown \cite{vogler_simulating_2013} to accurately describe thermally driven magnetization reversals of long-lived metastable nanostructures. 

In the FFS method, one defines a set of non-overlapping interfaces between the initial state, $A$, and the final state, $B$, of the transition. The interfaces are defined by requiring that an order parameter $\lambda(\boldsymbol{\hat{m}})$ takes particular values $\lambda_i$. Transition pathways connecting the initial with the final state are then constructed by considering short trajectory segments connecting adjacent interfaces. From the statistics of these trajectory segments one can then estimate the transition rate constant without the need to simulate the system during the long waiting times between subsequent events. The rate constant $k_{AB}$ for transitions from $A$ to $B$ is then given by
\begin{equation}
\label{eq:kab}
  k_{AB}=\frac{1}{\tau}={\Phi}_{A,1}\prod_{i=1}^{n-1}P\left ( \lambda_{i+1}\mid \lambda_{i} \right ),
\end{equation}
where $n$ is the total number of interfaces and ${\Phi}_{A,1}$ is the flux of trajectories coming from region $A$ and crossing interface $\lambda_{1}$. The order parameter values $\lambda_{0}$ and $\lambda_{n}$ define the boundaries of states $A$ and $B$, respectively. In the product on the right-hand side of Eq.~\ref{eq:kab},  the expression $P\left ( \lambda_{i+1}\mid \lambda_{i} \right )$ is the conditional probability of trajectories that have crossed interface $i$ to cross interface $i+1$ rather than returning to $A$. For further details of the FFS method, concerning both the general aspects as well as the actual implementation of the method in the case of thermally driven magnetization reversals of magnetic nanostructures, we refer to Ref.~\cite{vogler_simulating_2013}.

\begin{figure}[h!]
 \centering
  \includegraphics[width=0.6\columnwidth]{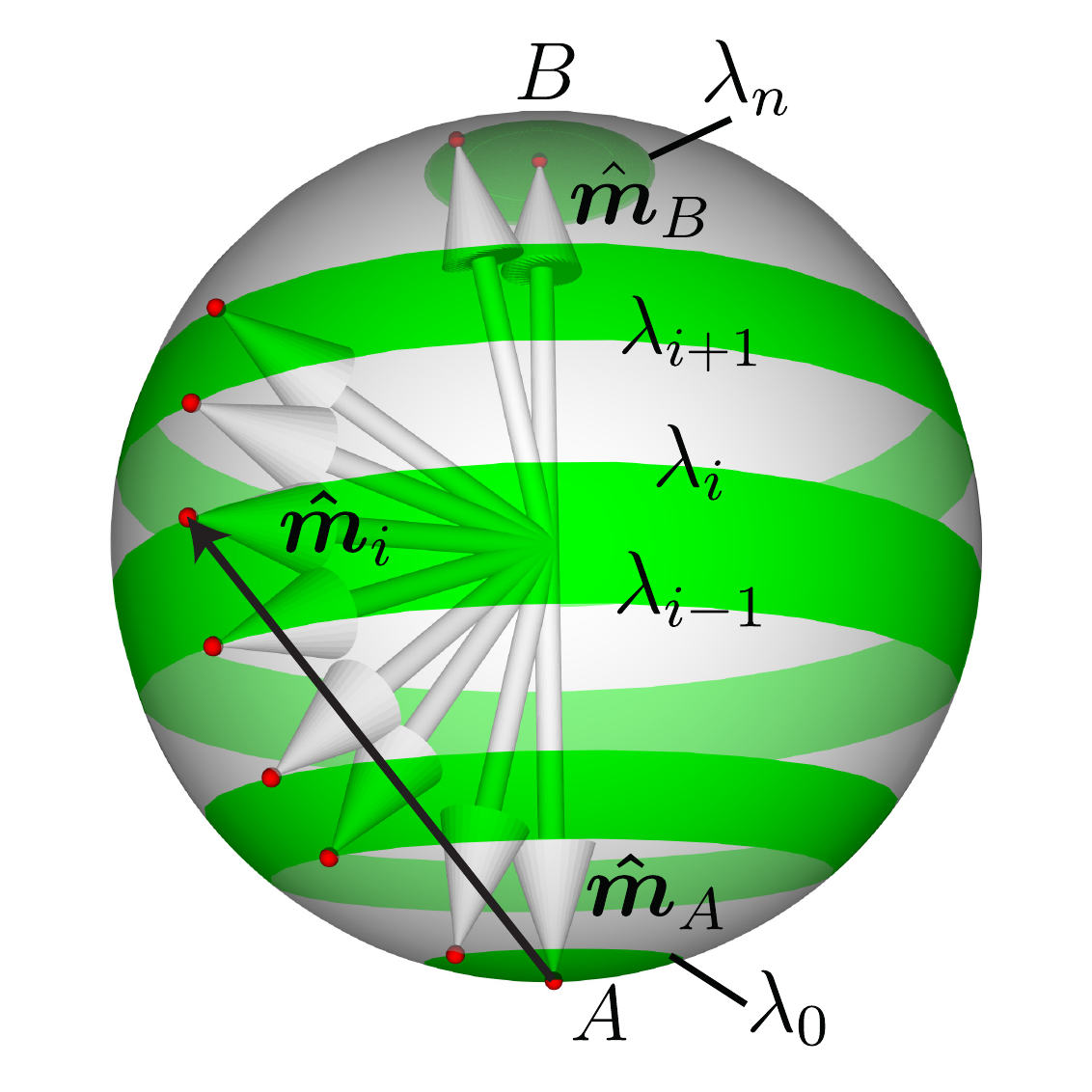}
  \caption{\small (Color online) Schematic illustration of the interfaces used in the FFS simulations. The two magnetic stable states $A$ and $B$ are defined by $\left
| \boldsymbol{\hat{m}}_{A}-\boldsymbol{\hat{m}}\right|\le \lambda_0$ and $\left|\boldsymbol{\hat{m}}_{B}-\boldsymbol{\hat{m}}\right| \le \lambda_n$, respectively. Here, the vertical bars denote the Euclidean norm.  The interfaces between $A$ and $B$ are defined by $\left
| \boldsymbol{\hat{m}}_{A}-\boldsymbol{\hat{m}}\right|= \lambda_i$ for $i=1, \ldots, n-1$ and correspond to the borders, labeled by $\lambda_{i}$, between the green and white areas on the surface of the unit sphere. The particular order parameter values $\lambda_i=\lambda(\boldsymbol{\hat{m}}_{i})$ are determined by evaluating the order parameter along a chain of states $\boldsymbol{\hat{m}}_i$ obtained from a NEB-calculation.
}
 \label{fig:interfaceDefinition}
\end{figure}

The stable states $A$ and $B$ as well as the interfaces between these states are defined in the following way. First, the minimum energy configurations $\boldsymbol{\hat{m}}_{A}$ and $\boldsymbol{\hat{m}}_{B}$ are determined by minimising the total energy $E_{\mathrm{tot}}$. Then, the NEB method is used to compute the minimum energy path between the energy minima, yielding a sequence of magnetic configurations $\boldsymbol{\hat{m}}_{i}$ with $0 \leq i \leq n$ along this path. To define the stable states $A$ and $B$ and the interfaces between them, we introduce the order parameter 
\begin{equation}
\label{eq:order_param}
 \lambda_{A}(\boldsymbol{\hat{m}})=\sqrt{\sum_{j=1}^{3K}\left ( m_{A,j}-m_j \right )^2},
\end{equation}
where $m_{A,j}$ and $m_j$ are the components of the minimum energy configuration $\boldsymbol{\hat{m}}_{A}$ and of the arbitrary magnetic configuration $\boldsymbol{\hat{m}}$, respectively. Hence, $\lambda_{A}(\boldsymbol{\hat{m}})$ is the Euclidean norm of $\boldsymbol{\hat{m}}_{A}-\boldsymbol{\hat{m}}$ measuring the distance of $\boldsymbol{\hat{m}}$ from $\boldsymbol{\hat{m}}_{A}$. The order parameter $\lambda_{B}(\boldsymbol{\hat{m}})$, quantifying the distance to the minimum energy configuration $\boldsymbol{\hat{m}}_{B}$, is defined analogously. Using these definitions, order parameter values can be assigned to any given magnetic configuration as illustrated in Fig.~\ref{fig:interfaceDefinition}. The interfaces are then defined by
\begin{equation}
\lambda_{A}(\boldsymbol{\hat{m}})=\lambda_i \quad\quad\quad 1 \leq i \leq n-1,
\end{equation}
where the order parameter values $\lambda_i$ are obtained by applying the order parameter $\lambda_A$ to the configurations resulting from the NEB-calculation, $ \lambda_i=\lambda_{A}(\boldsymbol{\hat{m_i}})$. The stables states $A$ and $B$ are defined similarly,
\begin{eqnarray}
 &&\boldsymbol{\hat{m}} \in A\quad\quad \text{if} \quad\quad \lambda_{A}(\boldsymbol{\hat{m}}) \leq \lambda_0 = \lambda_{A}(\boldsymbol{\hat{m}}_0)\nonumber \\
 &&\boldsymbol{\hat{m}} \in B\quad\quad \text{if} \quad\quad \lambda_{B}(\boldsymbol{\hat{m}}) \leq \lambda_n = \lambda_{B}(\boldsymbol{\hat{m}}_n).
\end{eqnarray}
Using these definitions one can easily check if the system resides in one of the stable states or on which side of a particular interface it is located. Note that the interface definitions given here differ somewhat from those of Ref.~\cite{vogler_simulating_2013}, in which the interfaces are constructed as hyperplanes. The new approach used here is computationally less expensive, but yields the same accuracy as the earlier method. In addition, the procedure for the optimisation of the interface locations \cite{borrero_optimizing_2008} converges considerably faster with the new interface definitions. 

Since in this work we aim at comparing the magnetic storage properties of single phase grains and graded media grains, we calculate their thermal stability in terms of the average retention time of the magnetic moments in one of the two equally probable energy minima. We examine one graded media and two single phase grains of different magnetic anisotropy. The same model geometry, an elongated prism with a pentagonal basal plane, is used for all investigated grains. The prism has a height of 20\,nm and the edge length of its basal plane is 3.53\,nm. Based on its uniaxial anisotropy, the preferred magnetization direction of the particle is pointing along the $z$-axis. Without the presence of an external magnetic field, the system has two stable states with all magnetic moments aligned in the $\pm z$-direction. The spatial discretization of the finite element calculations is the same for all models (54 nodes, 120 volume elements and 90 surface elements) as is the magnetic polarization $\mu_{0}M_{\mathrm{S}}=0.5\,\mathrm{T}$, the exchange constant $A=10^{-11}$\,J/m, and the damping constant $\alpha=0.02$. Material and simulation parameters only differ in the profile of the anisotropy constant $K_{1}$ along the grains. While a single phase grain consists of just one material, the graded media grain consists of several materials with a quadratically increasing anisotropy constant, as proposed by Suess et al. \cite{suess_thermal_2008}.

In order to compare the different grain models regarding their applicability in real memory devices, we compute their coercive fields by analyzing the hysteresis loops of the grains determined in  direct LLG simulations of the finite element model. The various coercive fields obtained from these calculations and the respective energy barriers of the magnetization reversals (obtained by the NEB method \cite{dittrich_path_2002,henkelman_climbing_2000}) are as follows. The graded media grain (GM), which consists of 8 segments with a quadratically increasing anisotropy constant, $K_{1}(z)=z^{2}\cdot4.57\times10^{21}\,\mathrm{J/m}^{5}$, has a coercive field of $\mu_{0}H_{\mathrm{C}}\,=\,1.11$\,T and an energy barrier of $\Delta E_{\mathrm{b}} = 53.2\,k_{\mathrm{B}}T$.  In addition, we investigate a soft and a hard magnetic single phase grain (SmSP and HmSP) with anisotropies of $K_{1}=1.9\times10^{5}\,\mathrm{J/m}^{3}$ and $K_{1}=6.61\times10^{5}\,\mathrm{J/m}^{3}$, respectively. The SmSP grain has the same coercive field ($\mu_{0}H_{\mathrm{C}}\,=\,1.11\, \mathrm{T}$) as the graded media grain, but a significantly lower energy barrier of $\Delta E_{\mathrm{b}} = 22.47\,k_{\mathrm{B}}T$. The HmSP has an energy barrier of $\Delta E_{\mathrm{b}}= 53.78\,k_{\mathrm{B}}T$, which is comparable to that of the graded media grain, but has a significantly larger coercive field of $\mu_{0}H_{\mathrm{C}}\,=\,3.3\, \mathrm{T}$. These properties are summarised in Tab.~\ref{tab:propertiesGrains}.

\setlength{\tabcolsep}{0.2cm}
\begin{table}[h!]
\centering
  \begin{tabular}{c | c D{.}{.}{2.2} D{.}{.}{2.2}}
    \toprule
    \toprule
    grain  & \multicolumn{1}{c}{$K_{1}$\,[MJ/m$^{3}$]} & \multicolumn{1}{c}{$\Delta E_{\mathrm{b}} / k_{\mathrm{B}}T$} & \multicolumn{1}{c}{$\mu_{0}H_{\mathrm{C}}$\,[T]} \\
    \midrule
    GM & $0.0-1.4$ & 53.62 & 1.11 \\
    SmSP & 0.19 & 22.47 & 1.11 \\
    HmS & 0.661  & 53.78 & 3.30 \\
    \bottomrule
    \bottomrule
  \end{tabular}
  \caption{\small Comparison of the most important properties of the three investigated grain models. $K_{1}$ is the anisotropy constant, $\Delta E_{\mathrm{b}}$ the energy barrier (in units of $k_{\mathrm{B}}T$) and $\mu_{0}H_{\mathrm{C}}$ the coercive field, obtained from simulated hysteresis loops.}
  \label{tab:propertiesGrains}
\end{table}

\begin{figure}[h!]
  \centering
  \includegraphics[width=1.0\columnwidth]{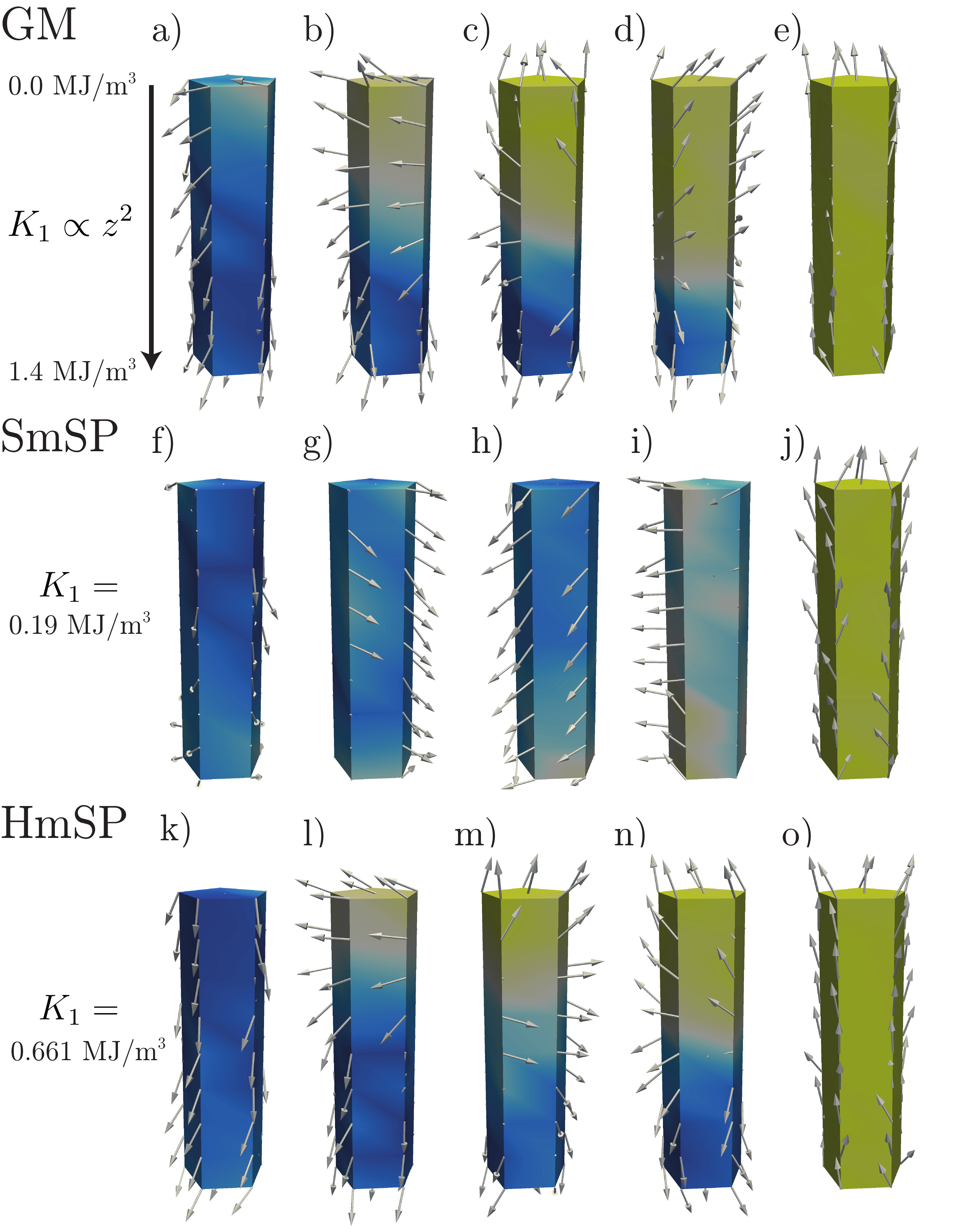}
  \caption{\small (Color online)  Exemplarily chosen magnetic configurations along the transition paths of thermally activated magnetization reversals of the investigated graded media (GM) grain, soft magnetic single phase (SmSP) grain and hard magnetic single phase (HmSP) grain, at a temperature of 300\,K. The GM grain has an anisotropy constant $K_{1}(z)\varpropto z^{2}$ increasing quadratically along the grain from zero to a maximum of $K_{1 \rm{,max}}=1.4\,\rm{MJ/m}^{3}$. The colors on the grains indicate the average normalized $z$-component of the magnetization.}
  \label{fig:FFS_transition_paths}
\end{figure}

Using FFS, we next compute the magnetization reversal rates for the three different grains. Due to the high barrier of the GM grain, 27 interfaces between the magnetic stable states with magnetization-down ($A$) and magnetization-up ($B$) are required to determine its thermal stability at a temperature of 300\,K. To estimate the accuracy of the computed rates, 10 FFS calculations with the same set of interfaces are performed. According to the optimization algorithm described in \cite{borrero_optimizing_2008}, the positions of the interfaces are then rearranged, in oder to provide a constant flux of partial trajectories through the interfaces for each of the 10 simulations ($15\,\% \leq \bar {p}_{i} \leq 21\,\%$). Figure~\ref{fig:FFS_transition_paths}a-e illustrates some representatively chosen magnetic configurations along a transition path of the reversal process of the GM grain. The magnetic moments start to reverse in the upper soft magnetic part of the grain, while precessing around its easy axis (Fig.~\ref{fig:FFS_transition_paths}a). A domain wall then forms (Fig.~\ref{fig:FFS_transition_paths}b) and moves downwards to the hard magnetic parts of the grain (Fig.~\ref{fig:FFS_transition_paths}c-d), until all magnetic moments are finally reversed (Fig.~\ref{fig:FFS_transition_paths}e) and the particle reaches the opposite (meta)stable state. After averaging over all 10 repeated FFS simulations the mean attempt frequency of the magnetic transition becomes $\bar{f_{0}}=8893.14$\,GHz, yielding an average grain stability of $\tau=718.07$ years. The standard deviation of $f_{0}$, estimated based on the repeated simulations, is $\sigma_{\mathrm{S}}=1997.95$\,GHz. One single FFS simulation requires an average of 21392.4 CPU hours. For a certainty interval of 99.7\,\% ($3\sigma_{\mathrm{S}}$), the GM grain is stable for at least for 412.77 years. The standard deviation $\sigma$ of the magnetization reversal rate computed in a FFS simulation can also be estimated using an analytical estimate derived by Allen et al. \cite{allen_forward_2006,allen_forward_2009}:
\begin{equation}
\label{eq:standard_deviation_single_FFS}
 \sigma = f_0 \sqrt{ \sum_{i=1}^{n-1} \frac{1-P\left ( \lambda_{i+1}\mid \lambda_{i} \right )}{P\left ( \lambda_{i+1}\mid \lambda_{i} \right ) r_i} }.
\end{equation}
Here, $n$ is the number of interfaces, $r_i$ is the number of trial trajectories starting at interface $i$ and $P\left ( \lambda_{i+1}\mid \lambda_{i} \right )$ are the transition probabilities between subsequent interfaces (see Eq.~\ref{eq:kab}). The corresponding standard deviations $\sigma$ obtained from the 10 simulations have a range of $602.80\mathrm{\,GHz} \leq \sigma \leq 1265.17\mathrm{\,GHz}$, which is in the order of magnitude of the standard deviation estimate based on the repeated FFS simulations, $\sigma_{\mathrm{S}}=1997.95$\,GHz. Hence, repeating FFS simulations with the same setup provides no additional information on the error of the rate, as the statistical properties of a single FFS simulation already allow for a reliable error estimation. For this reason, only single simulations are discussed from this point on.

Although the SmSP grain has the same coercivity as the GM grain, due to the lower energy barrier only 11 interfaces were required in the FFS simulation for the SmSP grain. Also in this case, the locations of the interfaces are optimized \cite{borrero_optimizing_2008} to produce a constant flux of partial trajectories ($17\,\% \leq p_{i} \leq 20\,\%$). 
A typical magnetization reversal path of the SmSP grain, obtained at a temperature of 300\,K, is shown in Fig.~\ref{fig:FFS_transition_paths}f-j. The difference to the transition path of the GM grain is obvious. As shown in Fig.~\ref{fig:FFS_transition_paths}f, the magnetic moments in all parts of the SmSP grain start to precess around the easy axis of the grain. In the course of the transition, depicted in Fig.~\ref{fig:FFS_transition_paths}g-j, the magnetic moments reverse almost homogeneously just like the moments in a sinle-domain particle. The FFS simulation of the SmSP grain yields an attempt frequency of $f_{0}=288.79$\,GHz corresponding to a thermal grain stability of $\tau=19.81$\,ms. The relative accuracy of the attempt frequency is the same as that of the GM grain.

Finally, we carry out a FFS simulation for the HmSP grain, which has the same energy barrier as the GM grain, but a much higher coercive field. The FFS simulation at a temperature of 300\,K requires  29 interfaces and the optimized \cite{borrero_optimizing_2008} partial flux of trajectories ranges from 18\,\% to 24\,\%. The whole FFS simulation requires 11255.27 CPU hours. The mechanism for magnetization reversal is similar to that of the GM grain. Inspecting a typical transition path in Fig.~\ref{fig:FFS_transition_paths}k-o, we notice that the reversal of the magnetic moments starts at one end of the grain (Fig.~\ref{fig:FFS_transition_paths}k). Then a domain wall forms (Fig.~\ref{fig:FFS_transition_paths}l) and propagates through the prism (Fig.~\ref{fig:FFS_transition_paths}m-n), until the transition is completed (Fig.~\ref{fig:FFS_transition_paths}o). Because of the symmetry of the system, it is equally probable for the domain wall to arise in the lower part of the particle as in the upper part, which is a main difference to the transition in the GM grain. The computed attempt frequency of the rare event is $f_{0}=6560.08$\,GHz, yielding a thermal stability of $\tau=1099.90$ years. 

\setlength{\tabcolsep}{0.2cm}
\begin{table}[h!]
  \centering
  \vspace{0.5cm}
  \begin{tabular}{c | D{.}{.}{2.3} D{.}{.}{4.4} r r}
    \toprule
    \toprule
    grain  &\multicolumn{1}{c}{$f_{0}$\,[GHz]} \hspace{0.2cm} & \multicolumn{1}{c}{$\tau$} & \multicolumn{1}{c}{$\sigma$\,[GHz]} & \multicolumn{1}{c}{CPU\,[h]} \\
    \midrule
    GM & 8893.14 \hspace{0.2cm} &718.07\,a & 825.90 & 21392.4 \\
    SmSP & 288.79 \hspace{0.2cm} & 19.81\,ms & 17.31 & 6727.1 \\
    HmS & 6560.08 \hspace{0.2cm} & 1099.90\,a & 613.57 & 11255.3 \\
    \bottomrule
    \bottomrule
  \end{tabular}
  \caption{\small FFS results for the graded media grain (GM), the soft magnetic single phase grain (SmSP), and the hard magnetic single phase grain (HmSP). Here, $f_{0}$ is the attempt frequency, $\tau$ the grain lifetime, and $\sigma$ is the standard deviation of $f_0$, computed according to Eq.~\ref{eq:standard_deviation_single_FFS}. The last column contains the CPU hours used in the FFS simulations. The GM grain results are averages over 10 repeated FFS simulations.}
  \label{tab:resultsCompare}
\end{table}

Tab.~\ref{tab:resultsCompare} summarizes the most important similarities and differences of the graded media grain and the two single phase grains with the same geometry. If one asks for a magnetic particle with low coercivity, one might expect that the GM grain or the SmSP grain to be a good choice (see Tab.~\ref{tab:propertiesGrains}). By looking at the thermal stabilities, however, it is obvious that the SmSP grain could never be used in a storage device, because it changes its magnetic state about 50 times per second (at 300\,K) on the average. In contrast, the GM grain has excellent thermal stability with a reversal rate of one per about 700 years. Comparing single phase with graded architecture (with the same energy barrier), the thermal stability of the HmSP grain is slightly higher, but its coercive field is much larger. Graded media grains can combine both low coercivity and high thermal stability and thus are able to overcome the superparamagnetic limit.

It has to be emphasized that the rate constants presented here are only qualitatively correct, meaning that the relative values of the thermal escape rates are accurate, but their absolute values depend on the spatial discretization length scale used in the finite element model of the particles. Since the strength of the thermal field $\boldsymbol{H}_{\rm th}$ in Eq.~\ref{eq:LLGThermal} is dependent on the discretization volume, the magnetization dynamics of the underlying grains are also mesh size dependent. In principle, an atomistic discretization should produce the correct dynamics and thus the correct escape rates. Nevertheless, the conclusions that 
\begin{itemize}
 \item a graded media grain with the same energy barrier as a single phase grain has a significantly lower coercive field
 \item a graded media grain with the same coercive field as a single phase grain has a significantly larger thermal stability 
\end{itemize}
remains valid. In summary, we have proven for the first time, using forward flux sampling, that the concept of magnetic grains with graded anisotropy provides very small nanostructures with high thermal stability and low coercivity. The results are obtained without any free parameters by directly integrating the underlying equation of motion (Eq.~\ref{eq:LLGThermal}). Due to the quadratically increasing anisotropy the grains combine the desired properties of both soft and hard magnetic parts, without suffering from their disadvantages.

The authors would like to thank the WWTF (Project MA09-029) and the FWF (Project SFB-ViCoM, F4112-N13) for financial support. The support from the CD-laboratory AMSEN (financed by the Austrian Federal Ministry of Economy, Family and Youth, the National Foundation for Research, Technology and Development) is acknowledged. The computational results presented have been achieved using the Vienna Scientific Cluster (VSC).


\begin{thebibliography}{15}%
\makeatletter
\providecommand \@ifxundefined [1]{%
 \@ifx{#1\undefined}
}%
\providecommand \@ifnum [1]{%
 \ifnum #1\expandafter \@firstoftwo
 \else \expandafter \@secondoftwo
 \fi
}%
\providecommand \@ifx [1]{%
 \ifx #1\expandafter \@firstoftwo
 \else \expandafter \@secondoftwo
 \fi
}%
\providecommand \natexlab [1]{#1}%
\providecommand \enquote  [1]{``#1''}%
\providecommand \bibnamefont  [1]{#1}%
\providecommand \bibfnamefont [1]{#1}%
\providecommand \citenamefont [1]{#1}%
\providecommand \href@noop [0]{\@secondoftwo}%
\providecommand \href [0]{\begingroup \@sanitize@url \@href}%
\providecommand \@href[1]{\@@startlink{#1}\@@href}%
\providecommand \@@href[1]{\endgroup#1\@@endlink}%
\providecommand \@sanitize@url [0]{\catcode `\\12\catcode `\$12\catcode
  `\&12\catcode `\#12\catcode `\^12\catcode `\_12\catcode `\%12\relax}%
\providecommand \@@startlink[1]{}%
\providecommand \@@endlink[0]{}%
\providecommand \url  [0]{\begingroup\@sanitize@url \@url }%
\providecommand \@url [1]{\endgroup\@href {#1}{\urlprefix }}%
\providecommand \urlprefix  [0]{URL }%
\providecommand \Eprint [0]{\href }%
\providecommand \doibase [0]{http://dx.doi.org/}%
\providecommand \selectlanguage [0]{\@gobble}%
\providecommand \bibinfo  [0]{\@secondoftwo}%
\providecommand \bibfield  [0]{\@secondoftwo}%
\providecommand \translation [1]{[#1]}%
\providecommand \BibitemOpen [0]{}%
\providecommand \bibitemStop [0]{}%
\providecommand \bibitemNoStop [0]{.\EOS\space}%
\providecommand \EOS [0]{\spacefactor3000\relax}%
\providecommand \BibitemShut  [1]{\csname bibitem#1\endcsname}%
\let\auto@bib@innerbib\@empty
\bibitem [{\citenamefont {Suess}(2006)}]{suess_multilayer_2006}%
  \BibitemOpen
  \bibfield  {author} {\bibinfo {author} {\bibfnamefont {D.}~\bibnamefont
  {Suess}},\ }\href@noop {} {\bibfield  {journal} {\bibinfo  {journal} {Appl.
  Phys. Lett.}\ }\textbf {\bibinfo {volume} {89}},\ \bibinfo {pages} {113105}
  (\bibinfo {year} {2006})}\BibitemShut {NoStop}%
\bibitem [{\citenamefont {Suess}\ \emph {et~al.}(2008)\citenamefont {Suess},
  \citenamefont {Fidler}, \citenamefont {Zimanyi}, \citenamefont {Schrefl},\
  and\ \citenamefont {Visscher}}]{suess_thermal_2008}%
  \BibitemOpen
  \bibfield  {author} {\bibinfo {author} {\bibfnamefont {D.}~\bibnamefont
  {Suess}}, \bibinfo {author} {\bibfnamefont {J.}~\bibnamefont {Fidler}},
  \bibinfo {author} {\bibfnamefont {G.}~\bibnamefont {Zimanyi}}, \bibinfo
  {author} {\bibfnamefont {T.}~\bibnamefont {Schrefl}}, \ and\ \bibinfo
  {author} {\bibfnamefont {P.}~\bibnamefont {Visscher}},\ }\href {\doibase
  doi:10.1063/1.2908052} {\bibfield  {journal} {\bibinfo  {journal} {Appl.
  Phys. Lett.}\ }\textbf {\bibinfo {volume} {92}},\ \bibinfo {pages} {173111}
  (\bibinfo {year} {2008})}\BibitemShut {NoStop}%
\bibitem [{\citenamefont {Henkelman}\ \emph {et~al.}(2000)\citenamefont
  {Henkelman}, \citenamefont {Uberuaga},\ and\ \citenamefont
  {J{\'o}nsson}}]{henkelman_climbing_2000}%
  \BibitemOpen
  \bibfield  {author} {\bibinfo {author} {\bibfnamefont {G.}~\bibnamefont
  {Henkelman}}, \bibinfo {author} {\bibfnamefont {B.~P.}\ \bibnamefont
  {Uberuaga}}, \ and\ \bibinfo {author} {\bibfnamefont {H.}~\bibnamefont
  {J{\'o}nsson}},\ }\href {\doibase 10.1063/1.1329672} {\bibfield  {journal}
  {\bibinfo  {journal} {J. Chem. Phys.}\ }\textbf {\bibinfo {volume} {113}},\
  \bibinfo {pages} {9901} (\bibinfo {year} {2000})}\BibitemShut {NoStop}%
\bibitem [{\citenamefont {Dittrich}\ \emph {et~al.}(2002)\citenamefont
  {Dittrich}, \citenamefont {Schrefl}, \citenamefont {Suess}, \citenamefont
  {Scholz}, \citenamefont {Forster},\ and\ \citenamefont
  {Fidler}}]{dittrich_path_2002}%
  \BibitemOpen
  \bibfield  {author} {\bibinfo {author} {\bibfnamefont {R.}~\bibnamefont
  {Dittrich}}, \bibinfo {author} {\bibfnamefont {T.}~\bibnamefont {Schrefl}},
  \bibinfo {author} {\bibfnamefont {D.}~\bibnamefont {Suess}}, \bibinfo
  {author} {\bibfnamefont {W.}~\bibnamefont {Scholz}}, \bibinfo {author}
  {\bibfnamefont {H.}~\bibnamefont {Forster}}, \ and\ \bibinfo {author}
  {\bibfnamefont {J.}~\bibnamefont {Fidler}},\ }\href {\doibase
  10.1016/S0304-8853(02)00388-8} {\bibfield  {journal} {\bibinfo  {journal} {J.
  Magn. Magn. Mater.}\ }\textbf {\bibinfo {volume} {250}},\ \bibinfo {pages}
  {12} (\bibinfo {year} {2002})}\BibitemShut {NoStop}%
\bibitem [{\citenamefont {Weinan}\ \emph {et~al.}(2002)\citenamefont {Weinan},
  \citenamefont {Ren},\ and\ \citenamefont {Vanden-Eijnden}}]{e_string_2002}%
  \BibitemOpen
  \bibfield  {author} {\bibinfo {author} {\bibfnamefont {E.}~\bibnamefont
  {Weinan}}, \bibinfo {author} {\bibfnamefont {W.}~\bibnamefont {Ren}}, \ and\
  \bibinfo {author} {\bibfnamefont {E.}~\bibnamefont {Vanden-Eijnden}},\ }\href
  {\doibase 10.1103/PhysRevB.66.052301} {\bibfield  {journal} {\bibinfo
  {journal} {Phys. Rev. B}\ }\textbf {\bibinfo {volume} {66}},\ \bibinfo
  {pages} {052301} (\bibinfo {year} {2002})}\BibitemShut {NoStop}%
\bibitem [{\citenamefont {Dobin}(2007)}]{dobin_attempt_frequency_2007}%
  \BibitemOpen
  \bibfield  {author} {\bibinfo {author} {\bibfnamefont {A.~Y.}\ \bibnamefont
  {Dobin}}\ }(\bibinfo {organization} {52nd Annual Conference on Magnetism and
  Magnetic Materials (MMM)},\ \bibinfo {year} {2007})\BibitemShut {NoStop}%
\bibitem [{\citenamefont {Dellago}\ \emph {et~al.}(1998)\citenamefont
  {Dellago}, \citenamefont {Bolhuis}, \citenamefont {Csajka},\ and\
  \citenamefont {Chandler}}]{dellago_transition_1998}%
  \BibitemOpen
  \bibfield  {author} {\bibinfo {author} {\bibfnamefont {C.}~\bibnamefont
  {Dellago}}, \bibinfo {author} {\bibfnamefont {P.~G.}\ \bibnamefont
  {Bolhuis}}, \bibinfo {author} {\bibfnamefont {F.~S.}\ \bibnamefont {Csajka}},
  \ and\ \bibinfo {author} {\bibfnamefont {D.}~\bibnamefont {Chandler}},\
  }\href {\doibase 10.1063/1.475562} {\bibfield  {journal} {\bibinfo  {journal}
  {J. Chem. Phys.}\ }\textbf {\bibinfo {volume} {108}},\ \bibinfo {pages}
  {1964} (\bibinfo {year} {1998})}\BibitemShut {NoStop}%
\bibitem [{\citenamefont {Bolhuis}\ \emph {et~al.}(2002)\citenamefont
  {Bolhuis}, \citenamefont {Chandler}, \citenamefont {Dellago},\ and\
  \citenamefont {Geissler}}]{bolhuis_transition_2002}%
  \BibitemOpen
  \bibfield  {author} {\bibinfo {author} {\bibfnamefont {P.~G.}\ \bibnamefont
  {Bolhuis}}, \bibinfo {author} {\bibfnamefont {D.}~\bibnamefont {Chandler}},
  \bibinfo {author} {\bibfnamefont {C.}~\bibnamefont {Dellago}}, \ and\
  \bibinfo {author} {\bibfnamefont {P.~L.}\ \bibnamefont {Geissler}},\ }\href
  {\doibase 10.1146/annurev.physchem.53.082301.113146} {\bibfield  {journal}
  {\bibinfo  {journal} {Annu. Rev. Phys. Chem.}\ }\textbf {\bibinfo {volume}
  {53}},\ \bibinfo {pages} {291} (\bibinfo {year} {2002})}\BibitemShut
  {NoStop}%
\bibitem [{\citenamefont {van Erp}\ \emph {et~al.}(2003)\citenamefont {van
  Erp}, \citenamefont {Moroni},\ and\ \citenamefont
  {Bolhuis}}]{van_erp_novel_2003}%
  \BibitemOpen
  \bibfield  {author} {\bibinfo {author} {\bibfnamefont {T.~S.}\ \bibnamefont
  {van Erp}}, \bibinfo {author} {\bibfnamefont {D.}~\bibnamefont {Moroni}}, \
  and\ \bibinfo {author} {\bibfnamefont {P.~G.}\ \bibnamefont {Bolhuis}},\
  }\href {\doibase doi:10.1063/1.1562614} {\bibfield  {journal} {\bibinfo
  {journal} {J. Chem. Phys.}\ }\textbf {\bibinfo {volume} {118}},\ \bibinfo
  {pages} {7762} (\bibinfo {year} {2003})}\BibitemShut {NoStop}%
\bibitem [{\citenamefont {Allen}\ \emph {et~al.}(2005)\citenamefont {Allen},
  \citenamefont {Warren},\ and\ \citenamefont {ten
  Wolde}}]{allen_sampling_2005}%
  \BibitemOpen
  \bibfield  {author} {\bibinfo {author} {\bibfnamefont {R.~J.}\ \bibnamefont
  {Allen}}, \bibinfo {author} {\bibfnamefont {P.~B.}\ \bibnamefont {Warren}}, \
  and\ \bibinfo {author} {\bibfnamefont {P.~R.}\ \bibnamefont {ten Wolde}},\
  }\href {\doibase 10.1103/PhysRevLett.94.018104} {\bibfield  {journal}
  {\bibinfo  {journal} {Phys. Rev. Lett.}\ }\textbf {\bibinfo {volume} {94}},\
  \bibinfo {pages} {018104} (\bibinfo {year} {2005})}\BibitemShut {NoStop}%
\bibitem [{\citenamefont {Allen}\ \emph
  {et~al.}(2006{\natexlab{a}})\citenamefont {Allen}, \citenamefont {Frenkel},\
  and\ \citenamefont {ten Wolde}}]{allen_simulating_2006}%
  \BibitemOpen
  \bibfield  {author} {\bibinfo {author} {\bibfnamefont {R.~J.}\ \bibnamefont
  {Allen}}, \bibinfo {author} {\bibfnamefont {D.}~\bibnamefont {Frenkel}}, \
  and\ \bibinfo {author} {\bibfnamefont {P.~R.}\ \bibnamefont {ten Wolde}},\
  }\href {\doibase doi:10.1063/1.2140273} {\bibfield  {journal} {\bibinfo
  {journal} {J. Chem. Phys.}\ }\textbf {\bibinfo {volume} {124}},\ \bibinfo
  {pages} {024102} (\bibinfo {year} {2006}{\natexlab{a}})}\BibitemShut
  {NoStop}%
\bibitem [{\citenamefont {Vogler}\ \emph {et~al.}(2013)\citenamefont {Vogler},
  \citenamefont {Bruckner}, \citenamefont {Bergmair}, \citenamefont {Huber},
  \citenamefont {Suess},\ and\ \citenamefont
  {Dellago}}]{vogler_simulating_2013}%
  \BibitemOpen
  \bibfield  {author} {\bibinfo {author} {\bibfnamefont {C.}~\bibnamefont
  {Vogler}}, \bibinfo {author} {\bibfnamefont {F.}~\bibnamefont {Bruckner}},
  \bibinfo {author} {\bibfnamefont {B.}~\bibnamefont {Bergmair}}, \bibinfo
  {author} {\bibfnamefont {T.}~\bibnamefont {Huber}}, \bibinfo {author}
  {\bibfnamefont {D.}~\bibnamefont {Suess}}, \ and\ \bibinfo {author}
  {\bibfnamefont {C.}~\bibnamefont {Dellago}},\ }\href {\doibase
  10.1103/PhysRevB.88.134409} {\bibfield  {journal} {\bibinfo  {journal}
  {Physical Review B}\ }\textbf {\bibinfo {volume} {88}},\ \bibinfo {pages}
  {134409} (\bibinfo {year} {2013})}\BibitemShut {NoStop}%
\bibitem [{\citenamefont {Borrero}\ and\ \citenamefont
  {Escobedo}(2008)}]{borrero_optimizing_2008}%
  \BibitemOpen
  \bibfield  {author} {\bibinfo {author} {\bibfnamefont {E.~E.}\ \bibnamefont
  {Borrero}}\ and\ \bibinfo {author} {\bibfnamefont {F.~A.}\ \bibnamefont
  {Escobedo}},\ }\href {\doibase doi:10.1063/1.2953325} {\bibfield  {journal}
  {\bibinfo  {journal} {J. Chem. Phys.}\ }\textbf {\bibinfo {volume} {129}},\
  \bibinfo {pages} {024115} (\bibinfo {year} {2008})}\BibitemShut {NoStop}%
\bibitem [{\citenamefont {Allen}\ \emph
  {et~al.}(2006{\natexlab{b}})\citenamefont {Allen}, \citenamefont {Frenkel},\
  and\ \citenamefont {ten Wolde}}]{allen_forward_2006}%
  \BibitemOpen
  \bibfield  {author} {\bibinfo {author} {\bibfnamefont {R.~J.}\ \bibnamefont
  {Allen}}, \bibinfo {author} {\bibfnamefont {D.}~\bibnamefont {Frenkel}}, \
  and\ \bibinfo {author} {\bibfnamefont {P.~R.}\ \bibnamefont {ten Wolde}},\
  }\href {\doibase 10.1063/1.2198827} {\bibfield  {journal} {\bibinfo
  {journal} {J. Chem. Phys.}\ }\textbf {\bibinfo {volume} {124}},\ \bibinfo
  {pages} {194111} (\bibinfo {year} {2006}{\natexlab{b}})}\BibitemShut
  {NoStop}%
\bibitem [{\citenamefont {Allen}\ \emph {et~al.}(2009)\citenamefont {Allen},
  \citenamefont {Valeriani},\ and\ \citenamefont {{Rein ten
  Wolde}}}]{allen_forward_2009}%
  \BibitemOpen
  \bibfield  {author} {\bibinfo {author} {\bibfnamefont {R.~J.}\ \bibnamefont
  {Allen}}, \bibinfo {author} {\bibfnamefont {C.}~\bibnamefont {Valeriani}}, \
  and\ \bibinfo {author} {\bibfnamefont {P.}~\bibnamefont {{Rein ten Wolde}}},\
  }\href {\doibase 10.1088/0953-8984/21/46/463102} {\bibfield  {journal}
  {\bibinfo  {journal} {J. Phys.-Condens. Mat.}\ }\textbf {\bibinfo {volume}
  {21}},\ \bibinfo {pages} {463102} (\bibinfo {year} {2009})}\BibitemShut
  {NoStop}%
\end{thebibliography}

%

\end{document}